\documentclass[%
 aps,
 amsmath,amssymb,
 reprint,%
]{revtex4-1}
\usepackage{graphicx}
\usepackage{dcolumn}
\usepackage{bm}

\usepackage[utf8]{inputenc}
\usepackage[T1]{fontenc}
\usepackage{mathptmx}

\begin{document}

\title{Enhancement of concentration of XeV and GeV centers in nanocrystalline diamond through $He^+$ irradiation}
\author{T. Chakraborty$^{1,2,*}$, K. J. Sankaran$^{1,2,\dagger}$, K. Srinivasu$^{3}$, R. Nongjai$^{4}$, K. Asokan$^{4}$, C. H. Chen$^{3}$, H. Niu$^{3}$, K. Haenen$^{1,2}$}

\affiliation{$^{1}$Institute for Materials Research (IMO), Hasselt University,3590 Diepenbeek, Belgium.}
\affiliation{$^{2}$IMOMEC, IMEC vzw, 3590 Diepenbeek, Belgium.}
\affiliation{$^{3}$Accelerator Laboratory, Nuclear Science and Technology Development Center, National Tsing Hua University, Hsinchu 30013, Taiwan.}
\affiliation{{$^{4}$Materials Science Group, Inter-University Accelerator Centre, New Delhi 110 067, India.}}

\altaffiliation[current affiliation: ]{QuTech and Kavli Institute of Nanoscience, Delft University of Technology, 2600 GA Delft, The Netherlands}
\altaffiliation[$^{\dagger}$current affiliation: ]{CSIR-Institute of Minerals and Materials Technology, Bhubaneswar 751013, India.}

\begin{abstract}
Atomic defect centers in diamond have been widely exploited
in numerous quantum applications like quantum information, sensing,
quantum photonics and so on. In this context, there is always a requirement to improve and optimize the preparation procedure to generate the defect centers in controlled fashion, and to explore new defect centers which can
have the potential to overcome the current technological challenges.
Through this letter we report enhancing the concentration of Ge and Xe vacancy centers in nanocrystalline diamond (NCD) by means of $He^{+}$ irradiation. We have demonstrated controlled growth of NCD by chemical vapor deposition (CVD) and implantation of Ge and Xe ions into the CVD-grown samples. NCDs were irradiated with $He^{+}$ ions and characterized through optical spectroscopy measurements. Recorded photoluminescence results revealed a clear signature of enhancement of the Xe-related and Ge vacancies in NCDs. 
\end{abstract}

\maketitle
In the last few decades, there have been notable advancements in the direction of synthesizing and growing diamond crystals in a controlled way which on one hand has boosted up the progress in material science and solid
state physics \citep{rudolph2014handbook,mildren2013optical}, and
on the other hand, has privileged the quantum revolution which exploits
the atomic defect centers in diamond \citep{Doherty:2013uq,Jelezko_review}. The Nitrogen vacancy (NV) center in diamond is established as a promising atomic defect center in diamond which has exhibited promising applications in quantum information processing \citep{Bernien_PRL,Neumann_Science}, magnetometry \citep{rondin2014magnetometry,wolf2015subpicotesla},
bio-sensing \citep{schirhagl2014nitrogen,haziza2017fluorescent},
thermometry \citep{wang2015high,neumann2013high} and so on. However,
the NV center has certain disadvantages: the photoluminescence (PL) signal
from the zero phonon line (ZPL) has only a contribution of $\sim4\%$
to the total spectrum where the rest of the signal is in phononic sidebands.
Moreover, the NV center possesses a permanent dipole moment which makes
it sensitive to external electrostatic fields. This is responsible
for the spectral diffusion of the NV optical transitions and hence
that affects the efficiency of generating single photons \citep{schmidgall2018frequency,acosta2012dynamic}. 

\begin{figure*}
\includegraphics[width=0.80\paperwidth]{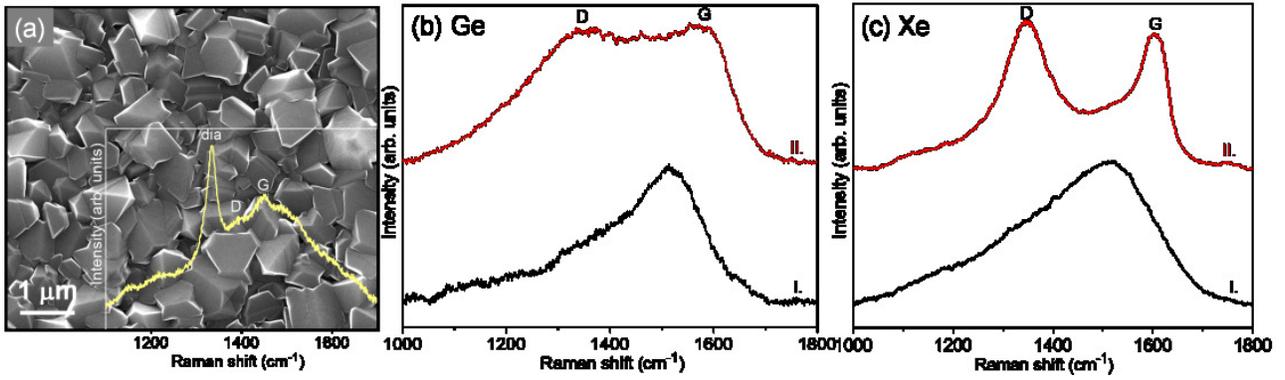}\caption{(a) SEM micrograph with inset of Raman spectrum of as-deposited NCD
films. (b) and (c) Raman spectra of (I) as-implanted and (II) He-ion
irradiated and annealed Ge and Xe-ion implanted NCD films. \label{fig: charcaterization}}
\end{figure*}

\begin{figure}

\includegraphics[width=1\columnwidth]{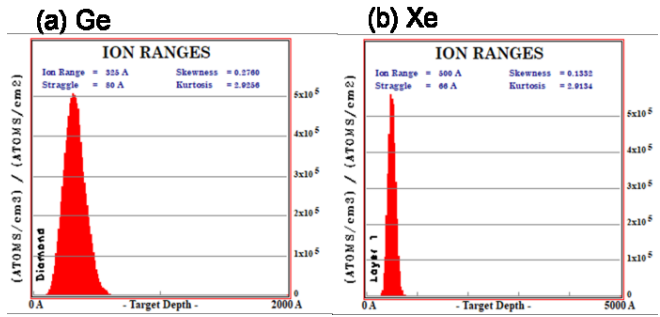}\caption{SRIM results of (a) Ge ion implantation and (b) SRIM results of Xe
ion implantation on NCD films. \label{fig:S1.-SRIM-results}}

\end{figure}

Elements of group IV of the periodic table, namely the SiV \citep{neu2011single,hepp2014electronic,becker2016ultrafast}
and GeV \citep{ekimov2015germanium,iwasaki2015germanium,zhou2018direct} centers in diamond have the potential to overcome the above-mentioned issues. The structure of these vacancy centers is inversion symmetric which makes them less susceptible to electrostatic fields. This results in a bright and narrowband emission into the ZPL and indistinguishability of the single photons emitted from multiple centers \citep{iwasaki2015germanium}. Specifically, the GeV center has certain advantages over the SiV center: because there is considerable probability that the excited state of the SiV center decays through certain nonradiative pathways, the SiV center has lesser quantum efficiency than that of the GeV center \citep{neu2013low}. These excellent spectral and spin properties make the GeV center a suitable system for quantum photonic and plasmonic applications \citep{bray2018single,siampour2018chip}. In this context, a constant experimental progress on the preparation procedure and better optimization of the growth parameters are necessary to grow GeV centers in diamond in a deterministic way, which is one of the main focuses of the present study.

In certain optical applications, it is advantageous to find out single photon sources which emit in the far-infrared region. For instance, in
fiber optic based communication the attenuation through silica glass fibers is considerably less in the infrared and far-infrared region in comparison to that in the visible range \citep{keiser2003optical}. The Xe-related vacancy (XeV) center in diamond, which exhibits a strong and a weak ZPL $\sim811$ nm and $\sim795$ nm, fulfills the above criterion. The XeV center is less investigated defect center in diamond \citep{sandstrom2018optical,zaitsev2006diamond,bergman2009photoluminescence}, whose exact crystallographic structure and spin properties are still to be explored. Considering its possible application in optical technology, we have focused on preparation and spectroscopic study of the XeV center. 

A widely used technique to create NV centers in diamond is through implanting high energy ($\sim$ MeV) nitrogen ions into the diamond substrate \citep{meijer2005generation,pezzagna2010creation} and subsequent annealing around 800 $^{\circ}$C \citep{chakraborty2019cvd} in ultra high vacuum (air pressure $\sim$ $10^{-10}$ mbar). However, such implantation technique requires a highly sophisticated ion accelerator, ion-focusing technique and supporting experimental set-up \citep{pezzagna2010creation,chakraborty2019cvd} that are expensive and is not easy to assemble, which hinders the easy production of NV centers. An alternative efficient procedure to make large scale production of vacancies is by irradiating the diamond substrate by low energy ions like protons and $He^{+}$ ions, which requires substantially less incident beam energy ($<$ 1 keV)\citep{fu2007characterization}. $He$ being a chemically inert gas, neutralization of the ions and embedment of $He$ atoms in diamond hardly affects the photo-physical properties of the NV center\citep{chang2008mass}. $He^{+}$ ions are proven to be more efficient than protons in terms of creating larger number of vacancies \citep{fu2007characterization}. With $He^{+}$ irradiation, the larger number of vacancies can be created with an implantation energy two orders of magnitude less than what is required for electron \citep{lawson1998existence,wee2007two,ziegler1985stopping} and $H^{+}$ implantation. $He^{+}$ beams with sufficient power required for creating vacancies can be generated using an RF ion source which can produce a current of two order magnitude higher than what can be generated for $H^{+}$ ions from an accelerator \citep{ziegler1985stopping}.

Earlier reports have demonstrated large scale preparation of stable NV centers in synthetic Nanocrystalline diamond (NCD) \citep{chang2008mass} with $He^{+}$ irradiation and the efficiency of the process has been investigated as a function of ion beam energy and dose \citep{kumar2018engineering}. We have investigated the effectiveness of $He^{+}$ irradiation on enhancing the concentration of the GeV and XeV centers through this work. We demonstrate controlled growth of NCD, creation of GeV and XeV centers through ion implantation, spectroscopic investigation of the created centers and the study of the effect of $He^{+}$ irradiation on the optical properties of the centers. We chose to create the GeV and XeV centers in NCDs as the preparation of color centers in NCD thin films is much cheaper compared to the single-crystalline diamond (SCD) layers. NCD films are grown on cheap materials like Si substrates whereas to grow an SCD layer, one needs SCD substrates which are much expensive. As a host of the color centers, NCDs are excellent candidate for applications in bioimaging, targeted drug delivery, sensing of temperature, spatial orientation, strain, pressure, magnetic and electric fields \citep{schirhagl2014nitrogen,merson2013nanodiamonds,chang2008mass}.

NCD films about 250 nm thick were deposited
onto Si (100) substrates using linear antenna microwave plasma enhanced
chemical vapor deposition. The deposition conditions of the
NCD films are reported elsewhere \citep{doi:10.1021/acs.cgd.7b00623,KUMAR2019107472}.
The pre-seeding process on the Si substrates was performed by spin coating a water-based state-of-the-art colloidal suspension of ultradispersed 6\textendash 7 nm sized detonation diamond particles. The scanning electron microscopy (FEI Quanta 200 FEG scanning electron microscope operated at 15 kV) image shown in Figure 1a illustrates that the as-deposited NCD films have uniform coverage of 1 $\mu$m sized faceted diamond grains. The inset of Fig. \ref{fig: charcaterization} (a) shows the Raman spectrum (a Horiba Jobin Yvon T64000 Raman spectrometer equipped with a BXFM Olympus 9/128 microscope in combination with a Horiba JY Symphony CCD detector with a 488 nm Lexell SHG laser) of as-deposited NCD films, revealing a sharp peak at 1332 cm$^{-1}$,
representing sp$^{3}$-diamond and the broad bands at \textasciitilde 1390 cm$^{-1}$ and \textasciitilde 1450 cm$^{-1}$ representing the disordered carbon (D) and graphitic (G) phases, respectively.

Ge and Xe ion implantations into NCD films were then carried out using the low energy ion beam facilities available at Inter University Accelerator Centre, (IUAC), New Delhi. We used 100 keV and 300 keV implantation energy for Ge and Xe ions, respectively, whereas the ion fluence was $1 \times{} 10^{15}$ ions/cm$^{2}$, for both ions. The penetration depth of implanted ions was obtained by SRIM. The Ge ions showed a penetration depth of 32 nm with a straggle of 8 nm for implantation energy of 100 keV {[}Fig. \ref{fig:S1.-SRIM-results} (a){]}, whereas penetration ~ 67 nm with a straggle of 11 nm was observed for the Xe-ions with an implantation energy of 300 keV {[}Fig.\ref{fig:S1.-SRIM-results} (b){]}. The Raman spectrum I in Fig. \ref{fig: charcaterization} (b) and (c) illustrates the amorphization induced in the NCD film implanted with the ion fluence of $1 \times{} 10^{15}$ ions/cm$^{2}$. The surface amorphization induced by implantation is converted into a graphitic phase because of the post-annealing process {[}spectrum II in Fig. \ref{fig: charcaterization} (b) and (c){]}. Such a phenomenon agrees with literature reports \citep{kalish1999doping,prawer1995ion}. The shift of the Raman peaks from spectra I to II is obvious due to the stress created by the implantated ions substituted at the carbon sites of the diamond lattice. In addition, the implantation and annealing induced amorphization and the size effect of newly formed nano-sized graphite are responsible for shifting the Raman peaks.
\begin{figure}
\includegraphics[width=1\columnwidth]{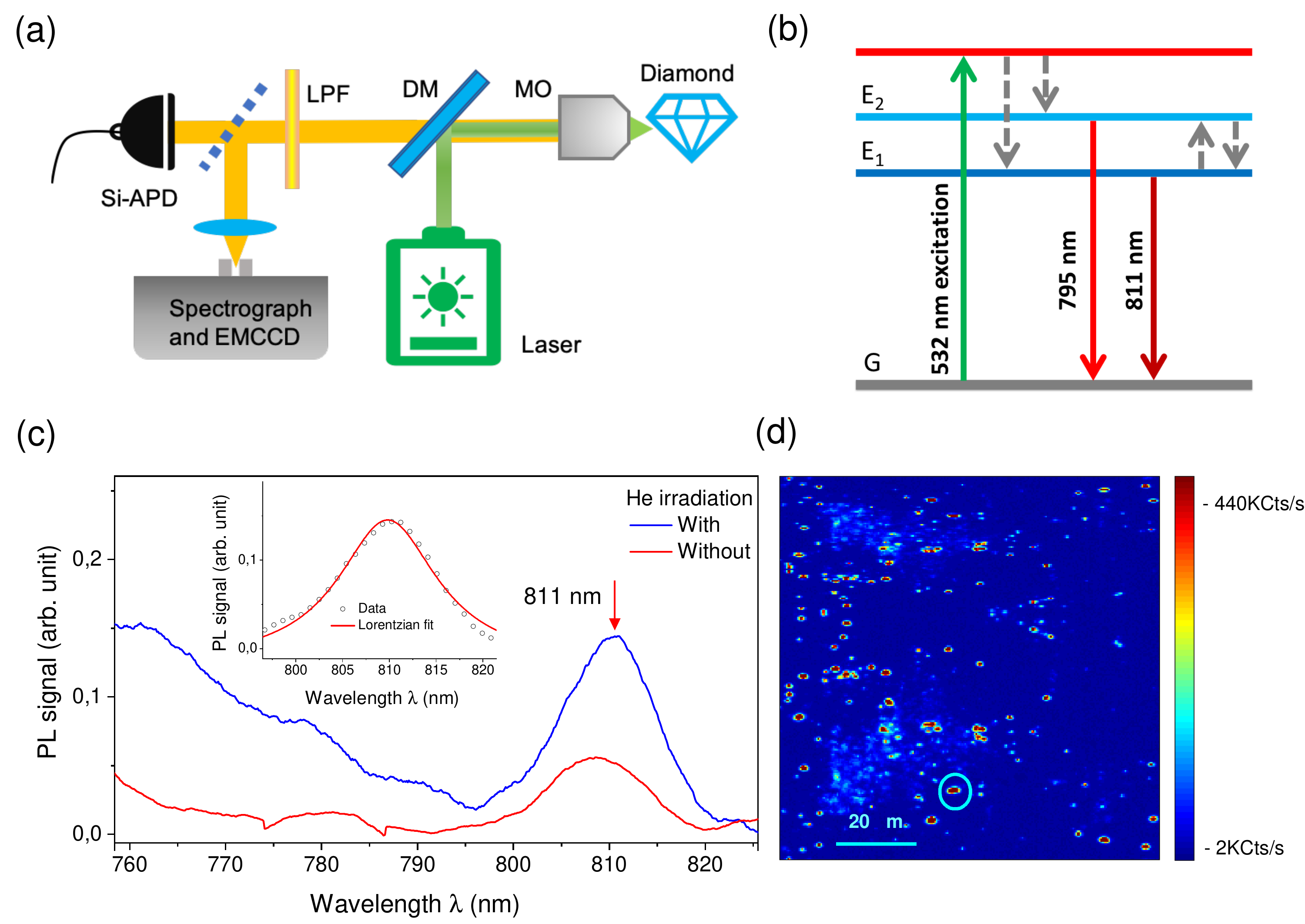}\caption{(a) The schematic representation of the home-built confocal microscope which is integrated with a spectrometer. The components of the set-up are mentioned in the main text. (b) Schematic diagram of the energy levels of the XeV center. 'G' represent the ground state, E$_{1}$ and E$_{2}$ are the excited states. The arrows depict the transitions between the energy levels as described in the text. (c) Photoluminescence spectra of the implanted XeV centers which were (blue curve) and were not (red curve) subjected to He irradiation. The inset shows Gaussian fit to the ZPL at 811 nm for the blue curve. (d) Intensity mapping image of the XeV centers in diamond.  \label{fig:XeV_PL}}
\end{figure}
\begin{figure}
\includegraphics[width=1\columnwidth]{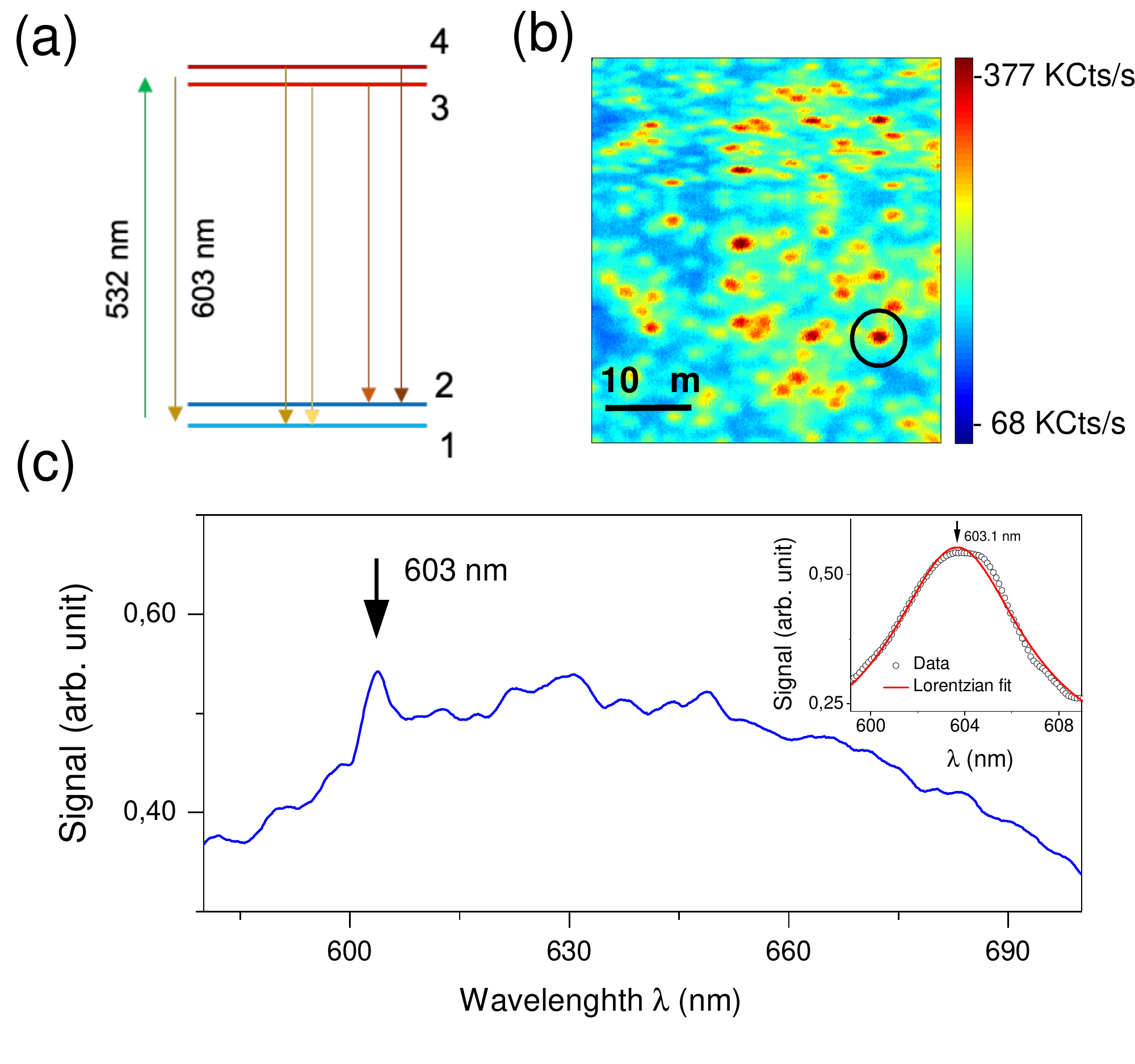}\caption{(a) Schematic diagram of the energy levels of the GeV center. (b) Intensity mapping image of the GeV centers measured using the confocal microscope. (c) Photoluminescence spectrum of the GeV centers marked by a circle in (b). The inset shows the background corrected data, with the ZPL and its fit to Gaussian peak function in an enlarged scale. \label{fig:Ge_optical_data}}
\end{figure}
After the ion-implantation, the samples were He-ion irradiated in a High Voltage Engineering Europa B.V. (HVE) particle accelerator. The energy of the $He^{+}$ ion beam was 100 keV with a fluence of $1 \times{} 10^{14}$ ions/cm$^{2}$. Subsequently, annealing was done at 900 $^{\circ}$ C for 2 h. 

To check the formation of the XeV and GeV centers, we performed optical characterization of the diamond samples in a home-built experimental setup where standard confocal microscopy was combined with optical spectroscopy. Our setup allows to capture photoluminescence (PL) mapping images on the XY plane of the diamond substrate as a function of depth and to perform optical spectroscopy on desired spots of the diamond substrate. 
Fig. \ref{fig:XeV_PL}(a) shows the schematic diagram
of the optical setup. The optical beam from a diode pumped 532 nm
laser is used for optical excitation. A high numerical aperture (NA=1.3) microscope objective (MO) focuses the laser beam onto the diamond sample and also collects the PL signal which is separated from the reflected laser by a dichroic mirror (DM) and a 550 nm long pass filter, and reaches the detection part of the set-up where a Silicon Avalanche Photodiode measures the signal. For creating mapping images of the sample, the MO is mounted on a nano-positioning piezo stage which has a traveling range of $100\mu m\times100\mu m$ in XY plane and $20\mu m$ along the Z axis. In order to measure spectra, we use a flip mirror after the DM to couple the signal to the input slit of a Andor Kymera 328i spectrograph combined with an Electron Multiplying charge-coupled device (EMCCD) camera, which has the sensitivity to detect optical signals at single photon level.

Fig. \ref{fig:XeV_PL}(b) exhibits a schematic representation
of the possible energy states of the XeV system which has been determined
through group theoretical calculations \citep{bergman2007polarization}.
It consists of a ground state G and two excited states E$_{1}$ and
E$_{2}$. The non-resonant 532 nm excitation triggers the transition
from the state G to A, a virtual absorption state. Subsequently, nonradiative transitions take place from the states A to the upper and lower excited states E$_{2}$ and E$_{1}$. The zero phonon lines $\thicksim$794 nm and $\thicksim$811 nm are associated with the transition E$_{2}$ to G and E$_{1}$to G, respectively. In order to examine the distribution of created XeV centers in the diamond nanocrystals through ion-implantation, we measured the PL signal and generated PL mapping images. The excitation power of the laser was 0.75 mW, which was measured right before the MO. The PL mapping images revealed bright spots distributed over certain locations of the diamond substrate.  To ensure the formation of XeV centers, we measured wavelength spectra of these bright spots.  Fig. \ref{fig:XeV_PL}(c)(the red curve) exhibits the 811 nm ZPL in such a spectrum measured for a certain bright spot. Keeping all the experimental parameters like laser power, measurement time, number of accumulation etc fixed for the spectroscopy measurements, we captured the spectra at different spots. Although signal intensity varied in the range $\thicksim$ $\pm$50$\%$, we were able to capture the $\thicksim$ 811 nm ZPL consistently at these spots. Thus the spectroscopy data ensures successful formation of the XeV center in ion-implanted samples. However, we could not observe the ZPL $\thicksim$794 nm between E$_{1}$ and G states. The 794 nm line being considerably weaker, its observation is hindered by the scattering and reflection of the signal through grain boundaries and rough surfaces of the nanodiamond crystals and the existence of a strong background possibly originated from the amorphous carbon and graphite. Moreover, there is a high probability that the signal gets absorbed by  $sp^2$ hybridized carbon atoms \citep{zaitsev2013optical}. Although the 811 nm signal has to encounter the above mentioned losses, the signal being comparatively stronger \citep{deshko2010spectroscopy}, we were able to capture it.  Next, after irradiating the sample with $He^{+}$ ions, we performed the optical measurements again keeping the experimental parameters the same. We captured PL data for a large number of positions throughout both the samples which was and was not irradiated by $He^{+}$ ions. We observed a significant enhancement in the average PL intensity for the $He^{+}$ irradiated sample in the mapping image and in the spectroscopy data accumulated through these measurements. Fig. \ref{fig:XeV_PL}(d) exhibits such a mapping image from an area of 100 $\mu$m $\times$ 100 $\mu$m. The optical spectrum was measured for the spot encircled in Fig. \ref{fig:XeV_PL}(d), and is shown by the blue curve in Fig. \ref{fig:XeV_PL}(d). The intensity of the spectra can be considered as approximately the mean intensity of several spectra measured at different spots. The spectrum evidently shows the ZPL peak $\thicksim$ 811 nm with almost 3 times enhanced intensity compared to the non-irradiated sample. The inset of Fig. \ref{fig:XeV_PL}(c) shows the fit of the XeV ZPL data to Lorentzian line-shape function and we observe a full width half maximum (FWHM) of 12.9 nm which is consistent with earlier reports\citep{sandstrom2018optical,zaitsev2006diamond}. In general, NCD films grown on non-diamond substrates lead to the growth of polycrystalline diamond films with intrinsic strain, which results in the broadening of ZPLs of color centers.

Incorporation of GeV centers in the diamond nanocrystals was
also investigated by combining confocal microscopy and optical spectroscopy using the optical set-up described above. Similar to SiV center \citep{hepp2014electronic}, the energy spectrum of the GeV center also consists of an excited state $^2{E}_g$ and ground state $^2{E}_u$, with both levels having two-fold degeneracy. One can assign level 1, 2 to the ground and 3, 4 to the excited states, respectively as shown in Fig. \ref{fig:Ge_optical_data}(a). Although at low temperature it is possible to spectrally resolve the 4 to 2, 4 to 1, 3 to 2 and 3 to 1 transitions, at RT one can only observe a single ZPL $\thicksim$602 nm. First, we performed measurements on the ion implanted sample. Although the mapping images showed bright spots distributed over the diamond substrate, the spots were not considerably bright compared to the background. The spectra measured at these spots showed either very weak or no peak close to 602 nm, which hardly confirms the creation of GeV centers. Next, to observe any change in the optical data, we measured the $He^{+}$ irradiated sample under the same experimental conditions. The PL mapping images depicted bright spots  with high contrast (SNR $>$ 5) being distributed inhomogeneously with varying concentration throughout the sample. Fig. \ref{fig:Ge_optical_data}(b) shows such a mapping image. To ensure the formation of the GeV centers, we performed optical spectroscopy measurements at different bright spots. The optical spectra from different positions consistently exhibit a distinct peak around 603 nm which can clearly be attributed to the room temperature (RT) ZPL of the GeV center. We measured all the spectra at same excitation laser power of 2 mW. The RT PL spectrum of the encircled spot shown in Fig. \ref{fig:Ge_optical_data}(b), is exhibited in Fig.
\ref{fig:Ge_optical_data}(c). Earlier observations reported that at RT the GeV ZPL appears at 602.5 nm \citep{ekimov2015germanium}, $\thicksim$602 nm \citep{iwasaki2015germanium}, 602.7 nm \citep{ralchenko2015observation} etc. However in our case we see the ZPL  $\thicksim$603 nm. ZPL being highly sensitive to local crystal strains \citep{batalov2009low}, the appearance of the peak position differs in different samples. The inset shows an enlarged view of the ZPL peak and its fit to Lorentzian line-shape, which has a FWHM of 4.5 nm which is consistent with earlier findings \citep{ralchenko2015observation,iwasaki2015germanium}. A significant enhancement in the PL intensity and consistence appearance of 603 nm peak at different positions throughout the sample suggest that $He^{+}$ irradiation and subsequent annealing has successfully created GeV centers in our diamond sample.
To conclude, we have demonstrated an engineered process of $He^{+}$ irradiation for creating XeV and GeV centers in nanocrystalline diamond and enhancing their concentration. $He^{+}$ irradiation and subsequent annealing at 900ºC were performed on the Ge and Xe ion implanted diamond samples. We captured the spectroscopic characteristics of both without and with $He^{+}$ irradiated samples in confocal microscope integrated with a spectrometer. We observed clear signature of increased PL signal of XeV and GeV centers for the irradiated samples, which evidently signifies an effect of $He^{+}$ irradiation. These results open the possibility for large scale creation of GeV and XeV centers in single crystal diamonds which can have potential application in quantum technology. Further optimization of the preparation and process parameters will allow us to maximize the production of GeV and XeV centers with reduced cost, which will favour their applicability in quantum photonics platforms.
\begin{acknowledgments} 
This work was financially supported by the Methusalem NANO network and the Research Foundation – Flanders (FWO) via SBO-project S004018N and the Special Research Fund (BOF) Postdoctoral Fellowship of KJS. We thank Milos Nesladek for his insightful suggestions. We also thank IUAC staff for their support in ion implantations.
\end{acknowledgments}

\bibliographystyle{apsrev4-1}
\bibliography{draft}

\end{document}